\documentclass[11pt]{article}
\usepackage{amsmath}  \usepackage{amsfonts}

\newcommand{\nag}{\nabla^{\Gamma}}
\newcommand{\naq}{\bar{\nabla}^{\Gamma}}
\newcommand{\nac}{\nabla}
\newcommand{\Ol}{O(\lambda)}

\newcommand{\nn}{\nonumber}
\newcommand{\be}{\begin{equation}}
\newcommand{\ee}{\end{equation}}

\newcommand{\va}{\varepsilon}
\newcommand{\ra}{\rightarrow}
\newcommand{\vf}{\varphi}
\newcommand{\Tr}{{\rm Tr}}

\newcommand{\p}{\partial}

\newcommand{\tF}{\tilde{F}} \newcommand{\tR}{\tilde{R}} \newcommand{\tS}{\tilde{S}}

\begin{document}
\title{\bf{Comparing metric and Palatini approaches to vector Horndeski theory}}
\author{E. A. Davydov\thanks{davydov@theor.jinr.ru}\\
\small{\it{Bogoliubov Laboratory of Theoretical Physics, JINR, }}\\
\small{\it{6 Joliot-Curie St, Dubna, Moscow region, 141980,
Russian Federation,}}\\
\small{and}
\\
\small{\it{Peoples' Friendship University of Russia (RUDN
University),}}\\
\small{\it{6 Miklukho-Maklaya St, Moscow, 117198, Russian
Federation}}}

\maketitle

\begin{abstract}
We compare cosmologic and spherically symmetric solutions to
metric and Palatini versions of vector Horndeski theory. It
appears that Palatini formulation of the theory admits more
degrees of freedom. Specifically, homogeneous isotropic
configuration is effectively bimetric, and static spherically
symmetric configuration contains non-metric connection. In
general, the exact solution in metric case coincides with the
approximative solution in Palatini case. The Palatini version of
the theory appears to be more complicated, but the resulting
non-linearity may be useful: we demonstrate that it allows the
specific cosmological solution to pass through singularity, which
is not possible in metric approach.
\end{abstract}

\section{Introduction}

The stages of accelerated expansion of our universe
were not predicted by Einstein gravity. This strongly motivates searching for new gravity theories. However, when we specify
the action of new theory, we immediately face the question: whether the
compatibility of connection with metric is an accidental feature
of Einstein gravity or a fundamental property of our universe?
Indeed, there are two main approaches to variation procedure of
the gravitational action, which are inequivalent. In
Palatini/first-order formalism the affine connection
$\Gamma^{\alpha}_{\rho\sigma}$ is treated as an independent
variable, while in metric/second-order formalism the connection is
restricted to be Levi-Civita connection of the metric
$g_{\rho\sigma}$: \be \nn
\Gamma^{\alpha}_{\rho\sigma}=\{^{\;\alpha}_{\rho\sigma}\}=\frac{g^{\alpha\lambda}}{2}
\left(\p_\rho g_{\lambda\sigma}+\p_\sigma
g_{\lambda\rho}-\p_\lambda g_{\rho\sigma}\right)\,.\ee Let us
denote, for brevity, the two formalisms as F1 and F2,
correspondingly.

Metric theory of gravity is present in any textbook on
gravitation. A detailed review of Palatini formulation of modified
gravity theories can be found in~\cite{Olmo:2011uz}. Worth to
mention, there exist a plethora of other options. One can consider
bimetric theory, in which connection is compatible with
independent second metric~\cite{Goenner:2010tr}, biconnection
theory with two affine connections~\cite{Tamanini:2012mi}, the
action may depend both on metric and affine geometrical
objects~\cite{Harko:2011nh} e.t.c. But those approaches are beyond
the scope of our consideration.

Einstein gravity appears to be physically equivalent in both
formalisms: even if connection is regarded as an independent
variable in Einstein-Hilbert action, the solution to corresponding
equation of motion is a connection, which admits the same geodesics
and Einstein equations as the Levi-Civita
connection~\cite{Dadhich:2010xa,Bernal:2016lhq}. There
are known few more theories which exhibit such
equivalence~\cite{Exirifard:2007da,Borunda:2008kf}, but for an
arbitrary gravity theory the equivalence is generally broken.
Identically looking actions provide different equations of motion
in F1 and F2 approaches~\cite{Iglesias:2007nv}. For instance,
$f(R)$ gravity contains an additional scalar degree of freedom,
compared to Einstein gravity. In F2 case, this degree of freedom
is dynamical and allows describing quint-essence and inflation. In
F1 this would be a non-dynamical degree of
freedom~\cite{Flanagan:2003rb}, which is similar to an effective
cosmological constant~\cite{Ferraris:1992dx,Sotiriou:2006qn}.

Consideration of independent connection may lead to ambiguities in
definition of geodesics, it admits traveling faster than light and
other potentially unpleasant phenomena, yet admissible at high
energies. On other hand, the artificial restriction on
connection may appear unphysical. Therefore both approaches to variation of
gravitational action are justified, and it is important to
establish an actual difference between F1 and F2 versions of
gravity theories actively studied nowadays.

In current research we would like to compare metric and Palatini
formulations of vector Horndeski theory. Horndeski models, both in
scalar~\cite{Horndeski:1974wa} and vector~\cite{Horndeski:1976gi}
case have attracted a lot of interest in recent years, because the
issues of inflation and/or dark energy can be resolved in a very
natural way within their
framework~\cite{Sushkov:2009hk,Starobinsky:2016kua,Davydov:2015epx}.
Although some of the models may be plagued by ghosts and
instabilities~\cite{BeltranJimenez:2017cbn}, their elegant
mathematical structure~\cite{Galileons1,Galileons2} motivates the
detailed study. The particular case of Horndeski prescription in
Palatini approach remains barely covered by the investigations,
though the theories with non-minimally coupled scalar field were
extensively explored recently in both metric and Palatini
formulations~\cite{Allemandi:2005qs,Harko:2010hw,Bauer:2008zj,Luo:2014eda}.
However, in Horndeski case the model with scalar field is quite
complicated, because it contains several distinct types of
coupling between matter and geometry. On the contrary, the vector
Horndeski model is unique in its simplicity, so we chose it as the
starting point for studying.

One can find several studies of
cosmological~\cite{Horndeski_Abelian1,Horndeski_Abelian2,Davydov:2015epx}
and spherically
symmetric~\cite{MuellerHoissen:1988bp,Balakin:2007am} solutions to
vector Horndeski theory in metric approach. But the Palatini
version of the theory remains unexplored, to the best of our
knowledge.Modified theories of gravitation with electromagnetic
fields were pretty well studied in the Palatini
approach~\cite{Teruel:2013rk,Baykal:2015paa,Olmo:2017qab}, but in
all the models examined, the interaction between matter and
gravity was realized at the level of effective scalars with the
lagrangians of the form:
$L=L(R,R_{\rho\sigma}R^{\rho\sigma},F_{\rho\sigma}F^{\rho\sigma},F_{\rho\sigma}\tF^{\rho\sigma})$.
And the vector Horndeski model exhibits much closer ties between
matter and geometry, implying contractions of curvature tensor
with the field tensors: \be L= {\tR}^{\alpha\beta\mu\nu}
F_{\alpha\beta} F_{\mu\nu} =
R_{\alpha\beta\mu\nu}\tF^{\alpha\beta}\tF^{\mu\nu} \label{03} \,.
\ee Here $R^{\alpha}_{\beta\mu\nu}$ is curvature tensor,
$F_{\mu\nu}$ --- field tensor, and their duals
${\tR}^{\alpha\beta\mu\nu}$, $\tF^{\alpha\beta}$ are obtained by
contractions with Levi-Civita tensor:
\begin{equation}\label{02}
\tR^{\alpha\beta\gamma\delta}=\frac14\,
\epsilon^{\alpha\beta\mu\nu}R_{\mu\nu\rho\sigma}\,\epsilon^{\rho\sigma\gamma\delta}\,,\quad
\tF^{\alpha\beta}=-\frac12\epsilon^{\alpha\beta\mu\nu}F_{\mu\nu}\,.
\end{equation}
Mention that compared to above, in literature the double dual of
Riemann tensor is often defined with minus sign.

\section{Non-minimal gravity in F1 and F2}

Despite focusing on vector Horndeski model, we start our research
from study of a gravity theory with non-minimally coupled matter
in general case. The reason is simple: we need to distinguish
peculiarities of Horndeski model and general features of a theory
with non-minimal coupling between matter and gravity. During this
general consideration we choose the coupling matter in a form of
scalar field, for simplicity. We will explain later that the
results will be also relevant to vector Horndeski theory.

So we start with the action\be S=\frac12\int\left[R-\lambda
L_c(R^{\alpha}_{\rho\sigma\nu},g_{\rho\sigma},\phi,\nabla_\mu\phi,\nabla_\rho\nabla_\sigma\phi)\right]\!\sqrt{-g}\,
d^4 x+\int L_m\sqrt{-g}\, d^4 x\,.\label{Lcoupl}\ee The minimally
coupled term $L_m$ represents some standard matter lagrangian
without curvature, connection and second-order derivatives. One
can naturally assume that at relatively low energies the
non-minimal coupling effects can be observed as small corrections
to Einstein gravity. The parameter of non-minimal coupling,
$\lambda$, may be used to establish energy scale at which the
coupling between matter and geometry becomes significant. Although
a particular physical interpretation of $\lambda$ depends on
specific lagrangian, we may formally assume that for a certain
class of theories the solutions can be expanded in powers of
$\lambda$, and the approximative solutions method can be applied.
Exact solutions in a theories with non-minimal coupling are rare,
so many results can be obtained with the aid of approximative
solutions.

\subsection{F1 case} Let us consider the Palatini approach, first.
Affine connection $\Gamma$ is then an independent variable, and
its difference from metric connection $\{\}$ is given by
distortion tensor: \be\label{dist0}
C^{\alpha}_{\cdot\;\rho\sigma}=\{^{\;\alpha}_{\rho\sigma}\}-\Gamma^{\alpha}_{\cdot\;\rho\sigma}\,.\ee
Throughout the article we will deal with symmetric connection
only:
$\Gamma^{\alpha}_{\cdot\;\rho\sigma}=\Gamma^{\alpha}_{\cdot\;\sigma\rho}$.
Though in $f(R,R_{\rho\sigma}R^{\rho\sigma})$ theories torsion
does not affect Einstein and field equations due to projective
invariance of scalar curvature~\cite{Sotiriou:2006qn}, in our case
there is a curvature tensor in the action. Hence setting torsion
to zero is a subtle question~\cite{Olmo:2013lta}. Nonetheless, we
let it vanish a priori, in order to simplify the calculations.

Covariant derivative with respect to metric connection will be
denoted as $\nac$, and covariant derivative with respect $\Gamma$
will be denoted as $\nag$.  The former preserves metric:
$\nac_{\alpha}g_{\rho\sigma}=0$. The latter, when acting on
metric, generates non-metricity tensor by the formula: \be
Q_{\alpha\rho\sigma}=-\nag_{\alpha}g_{\rho\sigma}\,;\quad
Q_\alpha\equiv\frac14
Q_{\alpha\rho\;\cdot}^{\;\cdot\;\cdot\;\rho}\,,\quad
\bar{Q}_\sigma \equiv
Q_{\;\cdot\;\alpha\sigma}^{\,\alpha\;\cdot\;\cdot}\,.\ee For
convenience of further use we've introduced also the traces of
non-metricity. The first of them, $Q_\alpha$, is often called Weyl
vector, and the factor $1/4$ is standard in its definition.
Mention that affine derivative of contravariant metric tensor is
equal to non-metricity with raised indices taken with positive
sign:
$\nag_{\alpha}g^{\rho\sigma}=Q_{\alpha}^{\;\cdot\;\rho\sigma}$.
The relation between distortion and non-metricity is following:
\be
C_{\alpha\rho\sigma}=-\frac12\left(Q_{\rho\alpha\sigma}+Q_{\sigma\alpha\rho}-Q_{\alpha\rho\sigma}\right)\,.\label{dist}\ee

The action~(\ref{Lcoupl}) now contains covariant derivatives with
respect to $\Gamma$, and Riemann tensor
$R^{\alpha}_{\rho\sigma\nu}$ depends solely on $\Gamma$: \be
S=\frac12\int\left[R(\Gamma,g)-\lambda
L_c(R^{\alpha}_{\rho\sigma\nu}(\Gamma),g_{\rho\sigma},\phi,\nag_\mu\phi,\nag_\rho\nag_\sigma\phi)\right]\!\sqrt{-g}\,
d^4 x+\int L_m\sqrt{-g}\, d^4 x\,.\label{LF1}\ee The derivation of
equations of motion usually requires a permutation of the
derivatives $\nag$. During this procedure, there appears a term
with Weyl vector due to presence of the factor $\sqrt{-g}$. For
example, in case of two arbitrary tensors $T$ and $K$ combined
into scalar density one has \be
\sqrt{-g}\,T^{\mu\alpha_{(n)}}_{\beta_{(k)}}\nag_\mu
K_{\alpha_{(n)}}^{\beta_{(k)}}=\p_\mu\left(\sqrt{-g}\,T^{\mu\alpha_{(n)}}_{\beta_{(k)}}
K_{\alpha_{(n)}}^{\beta_{(k)}}\right)-\sqrt{-g}\,K_{\alpha_{(n)}}^{\beta_{(k)}}(\nag_\mu-2Q_\mu)T^{\mu\alpha_{(n)}}_{\beta_{(k)}}\,.\ee
Thus it is useful to define the extended affine covariant
derivative: \be\label{naq_def} \naq_\nu \equiv \nag_\nu-2Q_\nu\,.
\ee The corresponding covariant divergence of a vector or
antisymmetric tensor coincides with metric covariant divergence:
\be \naq_\nu V^\nu=\nac_{\!\nu} V^\nu\,,\quad \naq_\nu
V^{[\nu\mu_1..\mu_n]}=\nac_{\!\nu} V^{[\nu\mu_1..\mu_n]}\,.
\label{naq}\ee Last property comes from the fact that both
connections $\Gamma^{\alpha}_{\beta\mu}$ and
$\{^{\;\alpha}_{\beta\mu}\}$ are symmetric and their contraction
in two indices with antisymmetric tensor identically vanishes.
Finally, let us introduce the two tensors in order to make the
formulas more compact: \be \label{MN}
M_{\alpha}^{\;\cdot\;\rho\sigma\nu}\equiv \frac{\p L_c}{\p
R^{\alpha}_{\rho\sigma\nu}}\,,\quad N^{\rho\sigma}\equiv\frac{\p
L_c}{\p\, \nag_{\rho}\nag_{\sigma}\phi}\,.\ee The first one is
antisymmetric in $\mu,\nu$, and the second one is symmetric.

Equation of motion for scalar field $\phi$ can be easily written
as \be -\lambda\naq_\rho\naq_\sigma N^{\rho\sigma}+\left[\nac_\mu
\frac{\p}{\p\p_\mu\phi}-\frac{\p}{\p\phi}\right]
\left(2L_m-\lambda L_c\right)=0\,.\label{Scaleq}\ee Mention that
affine and metric covariant divergences of the vector $\p L/\p_\mu
\phi$ coincide, as it follows from Eq.~(\ref{naq}).

Einstein equations in F1 look also quite simply:
\be\label{EinstF1}
G_{(\rho\sigma)}(\Gamma,g)=T^{(1)}_{\rho\sigma}(\Gamma,g)\,,\quad\mbox{where}\quad
T^{(1)}_{\rho\sigma}\equiv-\frac{\p}{\sqrt{-g}\p
g^{\rho\sigma}}\left[\left(2L_m-\lambda
L_c\right)\sqrt{-g}\right]\,.\ee Einstein tensor
$G_{\rho\sigma}(\Gamma,g)$ is not symmetric for arbitrary affine
connection $\Gamma$, and only its symmetric part joins the
equations.

In order to derive the connection equation one should vary the
lagrangian density from~(\ref{LF1}) in $\Gamma$: \be
\frac{\delta_{\Gamma}\left(\sqrt{-g}
L\right)}{\sqrt{-g}}=\frac12\left(\delta_{\alpha}^\sigma
g^{\rho\nu}-\lambda M_{\alpha}^{\rho\sigma\nu}\right)\delta_\Gamma
R^{\alpha}_{\rho\sigma\nu}(\Gamma)-\lambda
N^{\rho\sigma}\delta_\Gamma \nag_{\rho}\nag_{\sigma}\phi\,.\ee The
variation of curvature tensor with respect to connection reads as
\be \delta
R^{\alpha}_{\rho\sigma\nu}=\nag_\sigma\delta\Gamma^{\alpha}_{\rho\nu}-\nag_\nu\delta\Gamma^{\alpha}_{\rho\sigma}\,,\ee
while for second-order covariant derivative one has
\be\delta_\Gamma \nag_{\rho}\nag_{\sigma}\phi=-\nag_\alpha\phi\,
\delta\Gamma^{\alpha}_{\rho\sigma}\,. \ee Then the variation takes
the form
\begin{eqnarray}\frac{\delta_{\Gamma}\left(\sqrt{-g}
L\right)}{\sqrt{-g}}\,&=&-\left(\delta^{[\nu}_{\alpha}g^{\sigma]\rho}-\lambda
M_{\alpha}^{\rho\sigma\nu}\right)\nag_\nu\delta\Gamma^{\alpha}_{\rho\sigma}+
\frac{\lambda}{2} N^{\rho\sigma}\nag_\alpha\phi\, \delta\Gamma^{\alpha}_{\rho\sigma}\nn\\
&=&\left[\naq_\nu\left(\delta^{[\nu}_{\alpha}g^{\sigma]\rho}-\lambda
M_{\alpha}^{\rho\sigma\nu}\right)+\frac{\lambda}{2}
N^{\rho\sigma}\nag_\alpha\phi\right]\delta\Gamma^{\alpha}_{\rho\sigma}+\mbox{tot.
der.}\label{coneq0}\end{eqnarray} As a result, the equation on
symmetric affine connection reads as
\be\label{coneq1}\naq_\nu\left(\delta^{[\nu}_{\alpha}g^{\sigma]\rho}-\lambda
M_{\alpha}^{\rho\sigma\nu}\right)+\frac{\lambda}{2}
N^{\rho\sigma}\nag_\alpha\phi+(\rho\leftrightarrow\sigma)=0\,.\ee

Let us introduce hyperstress tensor: \be
\label{stress0}\Theta_{\alpha}^{\;\cdot\;\rho\sigma}\equiv
-\frac{2}{\sqrt{-g}}\frac{\delta S_c}{\delta
\Gamma^{\alpha}_{\rho\sigma}}=-2\lambda\naq_\nu
M_{\alpha}^{(\rho\sigma)\nu}+\lambda
N^{\rho\sigma}\naq_\alpha\phi\,. \ee It is the variational
derivative of non-minimal term in the action~(\ref{LF1}) with
respect to connection. Mention that the variation by connection of
the entire action is often called the connection tensor, and
variation of matter part of the action (coupled to connection via
covariant derivatives only) is called
hypermomentum~\cite{Hehl:1976kt}. Here we deal with the action
which is explicitly split onto Einstein-Hilbert and coupling
terms. The variation of EH term is trivial, while the variation of
coupling term is a matter of interest. However it is more than
just a material current coupled to connection, so it deserves an
individual name. By construction, hyperstress is symmetric in last
two indices, $\rho$ and $\sigma$.

The derivative $\naq$ acting on metric tensor spawns non-metricity
tensor and its traces. Therefore connection equation can be
rewritten as following: \be\label{Coneq}
Q^{\;\cdot\;\rho\sigma}_\alpha-2g^{\rho\sigma}Q_\alpha+2\delta_\alpha^{(\rho}Q^{\,\sigma)}-\delta_\alpha^{(\rho}\bar{Q}^{\,\sigma)}=-\Theta_{\alpha}^{\;\cdot\;\rho\sigma}\,.
\ee This is the differential equation for connection, but if $L_c$
is linear in curvature, it turns to algebraic one. It hardly can
be resolved in general, but one can use successive approximation
method, expanding the equation in powers of $\lambda$. For this,
let us expand the r.h.s. of connection equation~(\ref{Coneq}) in
powers of $\lambda$ up to $\Ol$ term. The result is the
hyperstress calculated on solutions to minimally coupled theory,
i.e. with $\lambda=0$, because the difference between solutions
with vanishing and non-vanishing $\lambda$ has $\Ol$ order, and
hyperstress already contains $\lambda$ as a factor.

The approximative solution for connection, valid in the order
$\Ol$, should therefore satisfy the equation\be\label{Coneq2}
Q^{\;\cdot\;\rho\sigma}_\alpha-2g^{\rho\sigma}Q_\alpha+2\delta_\alpha^{(\rho}Q^{\,\sigma)}-\delta_\alpha^{(\rho}\bar{Q}^{\,\sigma)}=
-\Theta_{\alpha}^{\;\cdot\;\rho\sigma}\,, \ee in which the r.h.s.
now is independent of affine connection and is known function
derived from solution to minimally coupled theory. It is
convenient also introduce the traces of hyperstress: \be
\Theta_\alpha\equiv
\Theta_{\alpha\rho\;\cdot}^{\;\cdot\;\cdot\;\rho}\,,\quad
\bar{\Theta}_\sigma \equiv
\Theta_{\;\cdot\;\alpha\sigma}^{\,\alpha\;\cdot\;\cdot}\,.\ee
Then, taking traces of the Eq.~(\ref{Coneq2}) and substituting it
back in the equation, after simple algebra one finds the
expression for non-metricity: \be
Q_{\alpha\rho\sigma}=\Theta_{\alpha\rho\sigma}-\frac23
g_{\alpha(\rho}\bar{\Theta}_{\sigma)}-
\left(\frac12\Theta_\alpha-\frac{1}{3}\bar{\Theta}_\alpha\right)g_{\rho\sigma}\,.
\label{nonmetricity}
 \ee
The distortion tensor can be obtained from~(\ref{dist}) with known
non-metricity: \be\label{dist2}
C_{\alpha\rho\sigma}=-\frac12\left[\Theta_{\rho\alpha\sigma}+\Theta_{\sigma\alpha\rho}-\Theta_{\alpha\rho\sigma}\right]-
\frac13 g_{\alpha(\rho}\bar{\Theta}_{\sigma)}+\frac12
g_{\alpha(\rho}\Theta_{\sigma)}-\frac{g_{\rho\sigma}}{4}\left(\Theta_\alpha-2\bar{\Theta}_\alpha\right)\,.
\ee We see that affine connection $\Gamma$ appears to be
incompatible with metric $g_{\rho\sigma}$, unless hyperstress is
vanishing.

\subsection{F2 case}
Now turn to metric formulation with the action\be
S=\frac12\int\left[R(g)-\lambda
L_c(R^{\alpha}_{\rho\sigma\nu}(g),g_{\rho\sigma},\phi,\nac_\mu\phi,\nac_\rho\nac_\sigma\phi)\right]\!\sqrt{-g}\,
d^4 x+\int L_m\sqrt{-g}\, d^4 x\,.\label{LF2}\ee The field
equation differs from F1 case~(\ref{Scaleq}) by replacement of the
covariant derivatives: \be -\lambda\nac_\rho\nac_\sigma
N^{\rho\sigma}+\left[\nac_\mu
\frac{\p}{\p\p_\mu\phi}-\frac{\p}{\p\phi}\right]
\left(2L_m-\lambda L_c\right)=0\,.\label{Scaleq2}\ee The terms
$N^{\mu\nu}$ and $L_c$ are also different in F1 and F2 cases,
because they depend on different connections. However, the
difference has the order $\Ol$, and $N^{\mu\nu}$ and $L_c$ enter
the equation already with the factor $\lambda$. Therefore the
equations for scalar field in F1 and F2 formulations are identical
in $\Ol$ order.

The derivation of Einstein equation in F2 case is a bit more
complicated, than in F1. Now the variation of connection and
curvature tensor should be expressed via metric variation: \be
% \nonumber to remove numbering (before each equation)
  \delta\Gamma^{\alpha}_{\rho\sigma}(g) = \frac{g^{\alpha\lambda}}{2}\left(\nac_\rho \delta
g_{\lambda\sigma}+\nac_\sigma\delta
g_{\lambda\rho}-\nac_\lambda\delta g_{\rho\sigma}\right)\,.
\label{convar} \ee As a result, in F2 version of Einstein
equations a new term, $T^{(2)}_{\rho\sigma}$, appears. In order to
calculate it, one should vary the coupling term with respect to
connection, and then take into account the dependence of
$\delta\Gamma$ on $\delta g$. So, on the first step we will get
nothing more than hyperstress~(\ref{stress0}), calculated on the
metric connection. After that, we should swap covariant
derivatives and apply the permutation of indices reflecting the
structure of the Eq.~(\ref{convar}). Combining all terms together
we obtain: \be\label{EinstF2}
G_{\rho\sigma}(g)=T^{(1)}_{\rho\sigma}(g)+T^{(2)}_{\rho\sigma}(g)\,,\quad\mbox{where}\quad
T^{(2)}_{\rho\sigma}(g)=-\frac12\nabla^\alpha\left(\Theta_{\rho\alpha\sigma}(g)+\Theta_{\sigma\alpha\rho}(g)-\Theta_{\alpha\rho\sigma}(g)\right)\,.
\ee The shape of term $T^{(1)}_{\rho\sigma}$ coincides with the
one derived in F1 case~(\ref{EinstF1}), but now it depends on
metric connection. Hyperstress in the expression for
$T^{(2)}_{\rho\sigma}$ also depends on metric connection
$\{^{\;\alpha}_{\rho\sigma}\}$. The fact that F1 and F2 versions
of Einstein equation differ by $T^{(2)}_{\rho\sigma}$ term was
already established in
literature~\cite{Borunda:2008kf,Cotsakis:1997cj}. And now we will
calculate the actual difference at the level of solutions,
expanding the equations in powers of $\lambda$.

\subsection{Comparing F1 and F2}
We have started with two actions~(\ref{LF1}),~(\ref{LF2}), which
just look similar but are actually different, and have applied two
different variational procedures. So no wonder that we eventually
obtained two different versions of Einstein equations.
Mathematically, F1 and F2 formulations are distinct, and that is a
well-known fact. Here we would like to accentuate that from
physical point of view the two formulations may appear less
distinguishable. When the modification to Einstein gravity is
considered as a correction at high energies, the corresponding
Observable phenomena should be barely visible at low energy scale
accessible to us. Those phenomena can be described by
approximative solutions. But what is the difference between
approximative solutions in F1 and F2?

Let us expand Einstein equations~(\ref{EinstF1},~\ref{EinstF2}) in
$\Ol$ order. All connection-dependent terms in right hand sides
already contain the factor $\lambda$, therefore the difference
between F1 and F2 connections will contribute in the r.h.s. only
into terms of the order $O(\lambda^2)$. Einstein tensor in l.h.s.
does not contain the factor $\lambda$, so the difference between
F1 and F2 versions of Einstein tensors, \be\delta
G_{\rho\sigma}=G_{(\rho\sigma)}(\Gamma,g)-G_{\rho\sigma}(g)\,,\ee
emerges already at $\Ol$ terms. In order to calculate $\delta
G_{\rho\sigma}$ in $\Ol$ order let us use the Palatini identity
\be \delta R_{(\mu\nu)}=\nac_\lambda
\delta\Gamma^{\lambda}_{\mu\nu}-\nac_{(\nu}\delta\Gamma^{\lambda}_{\mu)\lambda}\,,\ee
with difference between connections given by distortion
tensor~(\ref{dist2}):
\be\delta\Gamma^{\lambda}_{\mu\nu}=\Gamma^{\lambda}_{\mu\nu}-\{^{\;\alpha}_{\beta\mu}\}=-C^{\lambda}_{\mu\nu}\,.\ee
After short calculations one can see that \be \delta
G_{\rho\sigma}=\frac12
\nabla^\alpha\left(\Theta_{\rho\alpha\sigma}+\Theta_{\sigma\alpha\rho}-\Theta_{\alpha\rho\sigma}\right)+O(\lambda^2)\,.\label{dG}\ee
Shortly speaking, Einstein equations~(\ref{EinstF1}) and
(\ref{EinstF2}) in F1 and F2 cases look like \be\begin{split}
% \nonumber to remove numbering (before each equation)
  F1:&\quad G_{\rho\sigma}(g)-T^{(2)}_{\rho\sigma}(g)=T^{(1)}_{\rho\sigma}(g)+O(\lambda^2)\,,\nn \\
  F2:&\quad G_{\rho\sigma}(g)=T^{(1)}_{\rho\sigma}(g)+T^{(2)}_{\rho\sigma}(g)\,,\end{split}
\ee where all terms are evaluated for metric connection
$\{^{\;\alpha}_{\beta\mu}\}$.

The apparent equivalence of F1 and F2 versions of Einstein
equations in $\Ol$ order is not accidental, of cause. Both terms
$\delta G_{\rho\sigma}$ and $T^{(2)}_{\rho\sigma}$ reflect the
structure of solutions to connection equation. In F2 case the
equation on Levi-Civita connection is linear, while in F1 case the
equation on distortion tensor is linear only in $\Ol$ order. So it
appears that exact solutions to metric version of gravity
theory~(\ref{Lcoupl}) coincide with approximative $\Ol$ solutions
obtained in Palatini formulation of the theory.

\section{Vector Horndeski theory}

Let us now turn from general consideration to investigation of
particular theory. The covariant derivatives of vector and scalar
fields are different, so in general the results of previous
section can not be directly applied to theories with non-minimally
coupled vector field. However the lagrangian~(\ref{03}) does not
contain covariant derivatives of vector field due to our choice of
vanishing torsion. It contains only the first-order partial
derivatives, which are the same for vector and scalar fields.
Consequently, all calculations from previous section remain valid.

The lagrangian~(\ref{03}) for vector Horndeski model is given in
metric version. We suggest that it can be slightly modified to
suit better the F1 case. Gauge field comes as tensor
$\tF^{\alpha\beta}\tF^{\mu\nu}$ which is antisymmetric in first
and second pairs of indices and symmetric with respect to pair
exchange $(\alpha\beta)\leftrightarrow(\mu\nu)$. In metric case
the curvature tensor shares this symmetry, but not in Palatini
case. In latter case gauge field actually is coupled to
symmetrized part of curvature tensor (or its dual): \be
S_{\alpha\beta\mu\nu}\equiv
\frac12\left(g_{\lambda[\alpha}R^{\lambda}_{\;\beta]\mu\nu}+g_{\lambda[\mu}R^{\lambda}_{\;\nu]\alpha\beta}\right)\,,\quad
\tS^{\alpha\beta\gamma\delta}=\frac14\,
\epsilon^{\alpha\beta\mu\nu}S_{\mu\nu\rho\sigma}\,\epsilon^{\rho\sigma\gamma\delta}\,.\label{SS}
\ee The remaining components of curvature tensor do not interact
with $\tF^{\alpha\beta}\tF^{\mu\nu}$, because their contraction
identically vanishes. For metric curvature tensor one has
$S_{\alpha\beta\mu\nu}=R_{\alpha\beta\mu\nu}$, so the coupling in
the form $S\tF\tF=\tS F F$ correctly describes both F1 and F2
version of vector Horndeski model.

Finally, we would like to consider the action containing both
gravity-coupled and standard terms:
\begin{equation}\label{04}
S=\frac12\int\left[R-\Tr(\Phi^{\mu\nu}F_{\mu\nu})\right]\sqrt{-g}
d^4x\,.
\end{equation}
Here the ``induction'' tensor is introduced for convenience, it
absorbs minimally and non-minimally coupled field tensors: \be
\label{induction}
\Phi^{\mu\nu}=F^{\mu\nu}+\frac{\lambda}{2}\tS^{\mu\nu\lambda\tau}F_{\lambda\tau}\,.
\ee The trace operators appear because the vector field can be
non-Abelian with SU(2) gauge group. Then \be A_\mu=A^a_\mu
T_a\,,\quad F_{\mu\nu}=F_{\mu\nu}^a T_a=2\nabla_{[\mu}
A_{\nu]}-i\left[A_\mu,\,A_\nu\right]\,,\ee where SU(2) gauge group
generators are \be \left[T_a,T_b\right]=i\va_{ab}^{\;\;\;c}
T_c\,,\quad \Tr(T_a T_b)=\frac12\delta_{ab}\,. \ee In Maxwell case
trace operators should be replaced by factors $1\!/2$ in order to
get correct factors everywhere.

The field equation is the conservation law:
\begin{equation}\label{05}
\mbox{F1 and F2:}\quad D_\nu \Phi^{\mu\nu}
=0\,,\quad\mbox{where}\quad D_\nu\equiv\nac_\nu+[A_\nu,\cdot\,]\,.
\end{equation}
Covariant divergence here can be calculated with  metric
connection in both F2 and F1 approaches, because $\Phi^{\mu\nu}$
is antisymmetric, see Eq.~(\ref{naq}). Einstein equations look
quite different in two formulations:
\begin{eqnarray}
% \nonumber to remove numbering (before each equation)
   \mbox{F1}:&\quad G_{(\rho\sigma)} = &\Tr\left(
F_{(\rho\alpha}\left[\Phi_{\sigma)}^{\;\;\alpha}+F_{\sigma)}^{\;\;\alpha}\right]
    -\frac12 g_{\rho\sigma} F_{\alpha\beta}F^{\alpha\beta}\right)\,,\label{GeqF1}\\
  \mbox{F2}:&\quad G_{\rho\sigma}= &2\,\Tr\left(F_{(\rho\alpha}\Phi_{\sigma)}^{\;\;\alpha}
    -\frac14 g_{\rho\sigma} F_{\alpha\beta}\Phi^{\alpha\beta}\right)\nn\\
  && +\lambda\Tr\left(R^{\alpha\beta}\tF_{\alpha\rho}\tF_{\beta\sigma}
    +D^{\alpha}\tF_{\beta\rho}D^{\beta}\tF_{\alpha\sigma}+
    F^{\alpha\beta}\!\left[\tF_{\alpha\rho},\tF_{\beta\sigma}\right]\right)\,.\label{GeqF2}
\end{eqnarray}
The r.h.s. of connection equation~(\ref{Coneq}) now contains the
hyperstress\be
\label{stressF}\Theta_{\alpha}^{\;\cdot\;\rho\sigma}=-\lambda\naq_\nu\Tr\left(\tF_{\alpha}^{\;(\rho}\tF^{\sigma)\nu}\right)\,.\ee
The derivatives of hyperstress tensor arising in F2 version of
Einstein equations~(\ref{EinstF2}) generally contain higher order
derivatives of matter field. However, for Horndeski prescription
the terms with higher order derivatives are totally annihilated by
virtue of Bianchi identities for gauge field and curvature tensor.
The arbitrary couplings of the form $R F_{\mu\nu}F^{\mu\nu}$,
$R_{\mu\nu}F^{\mu\lambda}F^{\nu}_{\;\lambda}$, e.t.c. don't share
this property.

One can also check that in $\Ol$ order Einstein equations in F1
and F2 cases do coincide. For instance, the difference between
affine and metric Einstein tensors can be found from the
Eq.~(\ref{dG}) with hyperstress given above: \begin{eqnarray}
\delta G_{(\rho\sigma)}&=&\Tr\left(\frac12
g_{\rho\sigma}F_{\alpha\beta}\left[\Phi^{\alpha\beta}-F^{\alpha\beta}\right]-F_{(\rho\alpha}\left[\Phi_{\sigma)}^{\;\;\alpha}-F_{\sigma)}^{\;\;\alpha}\right]\right)\nn\\
&&-\lambda\Tr\left(R^{\alpha\beta}\tF_{\alpha\rho}\tF_{\beta\sigma}
+ D^\alpha\tF_{\beta\rho}D^\beta\tF_{\alpha\sigma}+
F^{\alpha\beta}[\tF_{\alpha\rho},\tF_{\beta\sigma}] \right)\,.
\end{eqnarray}
Combining it with F1 version of Einstein equation we obtain
precisely the F2 version.

Now let us compare the solutions to F1 and F2 versions of vector
Horndeski model beyond $\Ol$ order. The equations of motion are
quite complicated, so we will use suitable models mostly for
illustrative purposes. The detailed investigation of solutions in
each case is not the subject of the current consideration.

\subsection{Homogeneous isotropic model}
Unlike Maxwell case, the vector Horndeski model with SU(2)
Yang--Mills field admits homogeneous and isotropic cosmological
solutions, including inflationary ones~\cite{Davydov:2015epx}. So
here we consider the non-Abelian version of the action~(\ref{04})
The ansatz for metric in proper time gauge is standard:
\begin{equation}\label{4}
     ds^2=-dt^2+a^2\left[d\chi^2+\Sigma^{2}_{k}(\chi)
 (d\theta^2+\sin^2\theta d\varphi^2)\right]\,,
\end{equation}
where $\Sigma_{k}(\chi)=\{\sin\chi,\chi,\sinh\chi\}$ for the
closed, flat and open universe, labeled by $k=1,0,-1$,
correspondingly. For SU(2) Yang--Mills field the most general
cosmological ansatz preserving the isotropy and homogeneity of the
metrics can be written in terms of a single function $f(t)$ in all
three cases $k=0,\pm1$~\cite{BI}:
\begin{equation}\label{4a}
    A=f(t)T_\chi d\chi+\left[f(t)\Sigma_k
    T_\theta+(\Sigma_k'-1)T_\varphi\right]d\theta+ \left[f(t)\Sigma_k
    T_\varphi-(\Sigma_k'-1)T_\theta\right]\sin\theta d\varphi\,.
\end{equation}
Here the group generators, $T_a$, are the Pauli matrices
$\tau_b/(2i)$ contracted with spherical unit vectors
$n^b_{(\chi,\theta,\varphi)}$. Then the field tensor takes the
form~\cite{Galtsov:1991un}:
\begin{eqnarray}\label{4b}
F&=&\dot{f}\left(T_\chi\,dt\wedge d\chi
     + T_\theta \Sigma_k\,dt \wedge d\theta
     + T_\varphi \Sigma_k \sin \vartheta \,dt \wedge d\varphi \right) \nonumber \\
     &&+ \Sigma_k(f^2-k)\left(T_\varphi\,d\chi \wedge d\theta
     - T_\theta \sin \theta\,d\chi \wedge d\varphi
     + T_\chi \Sigma_k  \sin \theta\,d\theta \wedge d\varphi
     \right)\,.
\end{eqnarray}
Dot over the letter represents the time-derivative. Such
configuration (with $k=0$) in metric formulation was investigated
in~\cite{Davydov:2015epx,BeltranJimenez:2017cbn}, so here we may
focus on Palatini case.

It is easy to calculate hyperstress~(\ref{stressF}) in $\Ol$
order, since then we may use Levi-Civita connection of the
metric~(\ref{4}) instead of unknown affine connection. It appears
that all components of hyperstress in $\Ol$ order can be expressed
in terms of its trace: \be
\Theta_{\alpha\rho\sigma}=\frac13\left(\Theta_\alpha
g_{\rho\sigma}-\Theta_{(\rho} g_{\sigma)\alpha}\right)\,,\ee and
only the temporal component of the trace is non-vanishing: \be
\Theta_\alpha=3\lambda\left[\left(\frac{(f^2-k)^2}{a^4}-\frac{\dot{f}^2}{a^2}\right)H-\frac{1}{a^4}\frac{d}{dt}(f^2-k)^2\right]\delta_{\alpha
t}\,.\ee Here $H=\dot{a}\! /a$ is Hubble parameter, as usual. Then
the non-metricity tensor in $\Ol$ order can be found from the
Eq.~(\ref{nonmetricity}), and is equal to \be
Q_{\alpha\rho\sigma}=-\frac13\Theta_\alpha g_{\rho\sigma}\,.\ee It
is traceless, which happens only when covariant derivative is
compatible the conformally transformed metric $
h_{\rho\sigma}=e^{2\omega (t)}g_{\rho\sigma}$, where $\omega(t)$
is some unknown function.

One can check that the ansatz for connection taken in the form of
Levi-Civita connection compatible with metric $h_{\rho\sigma}$,
\be\label{Con_sphaleron}
\Gamma^{\alpha}_{\rho\sigma}=\frac{(h^{-1})^{\alpha\lambda}}{2}
\left(\p_\rho h_{\lambda\sigma}+\p_\sigma
h_{\lambda\rho}-\p_\lambda h_{\rho\sigma}\right)\,,\ee passes
through the full equations of motion (not restricted by $\Ol$
order), if $\dot{\omega}=-\Theta_t/6$. So, homogeneous isotropic
vector Horndeski model in Palatini approach is effectively a
bimetric theory. However the two metrics differ only by a
conformal factor, hence there is only one additional degree of
freedom compared to metric case.

The equations of motion are non-integrable and extremely
complicated. So we would like to simplify the model in order to
get the exact solutions, which can be easily analyzed. For this,
let us now turn to a so-called cosmological sphaleron solution.
The classical stress-energy tensor of minimally coupled theory
reads as: \be T_{t}^t=\rho\,,\quad
T_{i}^k=\frac{\rho}{3}\,\delta_i^k\,,\quad\mbox{where}\quad\rho=\frac32\left(\frac{\dot{f}^2}{a^2}+\frac{(f^2-k)^2}{a^4}\right)\,.\ee
In closed universe the gauge field potential $(f^2-1)^2/a^4$
acquires a double-valley form. The unstable static solution to
field equation, $f=0$, corresponds to the top of a barrier which
separates two distinct vacua states with $f=\pm 1$. It has
non-zero energy density $\rho=3/(2a^4)$ in Einstein gravity
theory. Hence one obtains the model with non-dynamical gauge field
and non-trivial stress-energy tensor which is very suitable for
investigation of complicated theory.

The equation $\dot{\omega}=-\Theta_t/6$ can be easily integrated,
since now $\Theta_t=3\lambda H/a^4$. Integrating it with natural
initial condition
$\Gamma^{\alpha}_{\rho\sigma}|_{\lambda=0}=\{^{\;\alpha}_{\rho\sigma}\}$
we find the exact solution to connection for cosmological
sphaleron model in F1 approach. It is the Levi-Civita connection
of conformally transformed metric \be
h_{\rho\sigma}=\exp\left(\frac{\lambda}{4a^4}\right)g_{\rho\sigma}\,.\label{metric_sphaleron}\ee
The factor $1/a^4$ in above formula is proportional to energy
density of Yang--Mills field. So when the field energy density
goes beyond the scale at which non-minimal coupling joins the
game, the difference between two metrics $g$ and $h$ starts
growing exponentially fast.

Now the Einstein equations~(\ref{GeqF1}--\ref{GeqF2}) are greatly
simplified and become integrable. We would like to present them in
a form of Friedmann equation:
\begin{eqnarray}
% \nonumber to remove numbering (before each equation)
   \mbox{F1}:&\quad H^2 = &\left(\frac{1}{2a^4}-\frac{1}{a^2}\right)\left(1-\frac{\lambda}{2
a^4}\right)^{-2}\,,\label{FriedmanF1}\\
  \mbox{F2}:&\quad H^2= &\left(\frac{1}{2a^4}-\frac{1}{a^2}\right)\left(1-\frac{\lambda}{
a^4}\right)^{-1}\,.\label{FriedmanF2}
\end{eqnarray}
Compared to Einstein gravity, the r.h.s. of the Friedmann
equations acquire additional factors, which differ starting from
$O(\lambda^2)$ order.

Let us briefly spell out the difference between Palatini and
metric versions of Horndeski theory in case of particular
cosmological sphaleron configuration. Non-minimal coupling in both
cases produces additional singularity when the Hubble parameter
diverges. In F1 case the singularity takes place at the point
$a^4=\lambda/2$, while in F2 --- at another point $a^4=\lambda$.
The behavior of solutions in vicinity of those points is distinct.
The r.h.s. of the Friedman equation~(\ref{FriedmanF2}) changes
sign at singularity, which is prohibited for l.h.s. Therefore the
F2 solution just stops there, which implies a so-called ``Big
Freeze'' singularity. In F1 case, the r.h.s. of the Friedmann
equation~(\ref{FriedmanF1}) remains positive after crossing the
critical point, and solutions can be analytically continued
through it. This picture resembles a phase transition in the
universe.

The behavior of F1 and F2 solutions in vicinity of initial
singularity, $a\ra 0$, is also different. In metric case the
Hubble parameter approaches constant, $H^2\ra-1/(2\lambda)$. Such
evolution of Hubble parameter takes place when the effective
equation of state corresponds to inflaton, $p=-\rho$. The Palatini
approach provides another asymptotic: $a\propto t^{-1/2}$, which
corresponds to phantom equation of state with $p=-\frac73\rho$.

\subsection{Static spherically-symmetric model}

We have found that for homogeneous isotropic configurations the
connection is compatible with some effective metric. However this
is not an inherent property of vector Horndeski model. Let us see
what happens in case of static spherically symmetric
configuration. The metric now reads as \be
ds^2=-w(r)dt^2+w(r)^{-1}dr^2+\rho(r)\left(d\theta^2+\sin^2\!\theta\,
d\varphi^2\right)\,, \ee and the ansatz for vector field is\be
A=f(r)\,dt+p\cos\theta\, d\vf\,,\ee which is often called the
scalar electrodynamics. There is no need in non-Abelian
configuration here, so we better consider the Maxwell field for
simplicity. The solutions to metric version of such model can be
found in literature~\cite{MuellerHoissen:1988bp,Balakin:2007am},
so we proceed with investigation of Palatini case.

The expression for hyperstress in $\Ol$ order is not so simple,
and does not allow to guess the ansatz for connection. However the
connection equation~(\ref{Coneq}) is linear in connection, because
the action of Horndeski theory~(\ref{04}) is linear in curvature.
The system of linear algebraic equations can be easily resolved
with a computer, so we can only present the results. In purely
electric case ($p=0$) one has the following non-vanishing
components of connection: \be
\begin{split}
&\Gamma^{t}_{tr}=\frac{w'}{2w}+\frac{\rho'
f'^2}{2\rho\mu^2}\left(\frac23+\lambda\frac{f'^2}{24}\right)\,,\quad
\Gamma^{r}_{tt}=w^2\left[\frac{w'}{2w}+\lambda\frac{\rho'
f'^2}{2\rho}\left(1+\lambda\frac{f'^2}{24}\right)\right]\,,\\
&\Gamma^{r}_{rr}=-\frac{w'}{2w}+\lambda\frac{\rho'
f'^2}{2\rho}\left(\frac13+\lambda\frac{f'^2}{24}\right)\,,\quad
\Gamma^{\theta}_{r\theta}=\Gamma^{\vf}_{r\vf}=\frac{\rho'}{\rho}\left(1+\lambda\frac{f'^2}{12}\right)\,,\\
&\Gamma^{r}_{\theta\theta}=-w\rho\rho'\,,\quad
\Gamma^{r}_{\vf\vf}=\Gamma^{r}_{\theta\theta}\sin^2\theta\,,\quad
\Gamma^{\theta}_{\vf\vf}=-\sin\theta\cos\theta\,,\quad\Gamma^{\vf}_{\theta\vf}=\frac{\cos\theta}{\sin\theta}\,.
\end{split}
\ee The solution to magnetic case ($q=0$) reads as \be
\begin{split}
&\Gamma^{t}_{tr}=-\Gamma^{r}_{rr}=\frac{w'}{2w}+\lambda\frac{\rho'
p^2}{6\rho^5}\left(1+\lambda\frac{p^2}{6\rho^4}\right)^{-1}\,,\quad
\Gamma^{r}_{tt}=w^2\left[\frac{w'}{2w}+\lambda\frac{\rho'
p^2}{2\rho^5}\left(1+\lambda\frac{p^2}{6\rho^4}\right)^{-1}\right]\,,\\
&\Gamma^{\theta}_{r\theta}=\Gamma^{\vf}_{r\vf}=
\frac{\rho'}{\rho}\left(1-\lambda\frac{p^2}{6\rho^4}\right)\left(1+\lambda\frac{p^2}{6\rho^4}\right)^{-1}\,,\quad
\Gamma^{r}_{\theta\theta}=-w\rho\rho'\left(1-\lambda\frac{p^2}{2\rho^4}\right)\,,\\
&\Gamma^{r}_{\vf\vf}=\Gamma^{r}_{\theta\theta}\sin^2\theta\,,\quad
\Gamma^{\theta}_{\vf\vf}=-\sin\theta\cos\theta\,,\quad\Gamma^{\vf}_{\theta\vf}=\frac{\cos\theta}{\sin\theta}\,.
\end{split}
\ee The solution which incorporates both $q$, $p$ non-zero is a
quite complicated combination of the above solutions, and there is
no need presenting it.

In electric case the connection is parameterized by five
nontrivial functions, while in magnetic case there are only four.
However, neither of these connections is Levi-Civita one. One can
easily show this considering just $\Ol$ order. If connection is
non-metric in that order, it is non-metric in general.

Suppose that connection is compatible with some metric
$h_{\rho\sigma}$, so that $\nag_\alpha h_{\rho\sigma}=0$. Let us
then calculate the difference between the two metrics,  $\delta
h_{\rho\sigma}\equiv (h_{\rho\sigma}-g_{\rho\sigma})$ up to $\Ol$
order. By definition, \be \nag_{\alpha}\delta
h_{\rho\sigma}=-\nag_{\alpha}g_{\rho\sigma}=Q_{\alpha\rho\sigma}\,.\ee
From other hand, $\nag_{\alpha}\delta
h_{\rho\sigma}=\nac_{\!\alpha}\delta h_{\rho\sigma}+O(\lambda^2)$.
Consequently, in $\Ol$ order one has \be \nac_{\!\alpha}\,\delta
h_{\rho\sigma}=Q_{\alpha\rho\sigma}\,,\label{dh1}\ee where
non-metricity tensor is given by the Eq.~(\ref{nonmetricity}). The
commutator of two covariant derivatives, $[\nac_\nu,\nac_\alpha]$,
generates algebraic equations on $\delta h_{\rho\sigma}$: \be
\delta
h_{\lambda(\rho}R^{\lambda}_{\;\sigma)\nu\alpha}=\nabla_{[\nu}Q_{\alpha]\rho\sigma}\,.\label{dh2}\ee
Here the Riemann tensor and covariant derivatives of non-metricity
can be calculated on solutions to non-coupled theory (with
$\lambda=0$). In our case the corresponding solution is
Reissner-Nordstr\"{o}m metric: \be f=\frac{q}{r}\,,\quad
\rho=r\,,\quad w=1-\frac{M}{r}+\frac{p^2+q^2}{4r^2}\,.\ee Then it
is not difficult to find the Riemann tensor and
hyperstress~(\ref{stress0}). Non-metricity will be given by the
Eq.~(\ref{nonmetricity}), and all covariant derivatives should be
taken with metric connection.

The number of equations in~(\ref{dh2}) exceeds the number of
independent variables $\delta h_{\rho\sigma}$, so the existence of
solution is not guaranteed, in general.  The system of linear
algebraic equations on ten functions $\delta h_{\rho\sigma}$ with
known coefficients can be easily investigated. In case of
considered spherically-symmetric configuration there are no
solutions to the Eq.~(\ref{dh2}). The Horndeski prescription ruins
the metricity of connection even in a case of
spherically-symmetric configuration. Mention that in Palatini
version of modified gravities with minimally coupled Maxwell field
the connection remains
metric-compatible~\cite{Teruel:2013rk,Olmo:2012nx}.

Unfortunately, now there are no such simple exact solutions like
those obtained in cosmological case. The equations on metric
functions are very complicated. Since our goal was to establish
the fundamental difference between F1 and F2 approaches, we will
not go into detailed investigation of the solutions. The main
result is that connection is non-metric in Palatini formulation.

\section{Conclusion}

It is a well-known fact that metric and Palatini versions of
modified gravity theories are mathematically distinct, but the
actual difference for specific theories remains unknown. Here we
have compared the most common solutions (homogeneous isotropic and
static with spherical symmetry) to vector Horndeski model, which
were derived in two formalisms. Though that is not a general
consideration, it provides good practical insight on the
difference between metric and Palatini versions of the vector
Horndeski theory.

It appears that in Palatini case there are more degrees of
freedom, because independent connection is not compatible with the
metric. The connection may be compatible with another metric, or,
most probably, it will be non-metric. The connection equation for
particular Horndeski prescription is linear and algebraic. One may
solve it and substitute the solution into Einstein equations.
After carrying this procedure, one can find that Palatini version
of Einstein equations incorporates metric Einstein equations as
$\Ol$ term. However there will be also the terms with higher
orders in $\lambda$, which makes the equations much more
complicated. But it does not make them worse. For example, in
cosmological sphaleron model the non-linearity allows to
analytically continue the solution through the certain
singularity, which is not possible in metric case. Thus, both
approaches have their advantages and disadvantages and deserve
equal consideration.

\section{Acknowledgment}
The author thanks prof. D.~V. Gal'tsov for his great contribution
to the work. The publication was financially supported by the
Ministry of Education and Science of the Russian Federation (the
Agreement number 02.a03.21.0008) and by the Russian Foundation for
Fundamental Research under grant 17-02-01299.

%========================================================================================================
%========================================================================================================
%========================================================================================================
% END OF DOCUMENT
%========================================================================================================
%========================================================================================================
%========================================================================================================
\end{document}